\documentclass[runningheads]{llncs}
\usepackage[T1]{fontenc}
\usepackage{amsmath,amssymb,amsfonts}
\usepackage{graphicx}
\usepackage{svg}
\usepackage{caption}
\usepackage{wrapfig}
\usepackage{subfigure}
\usepackage{pgfplots}
\usetikzlibrary{intersections}
\usepackage{bm}
\usetikzlibrary{positioning}
\usepgfplotslibrary{fillbetween}
\pgfplotsset{compat=1.5}
\usetikzlibrary{pgfplots.groupplots}
\usetikzlibrary{shapes.geometric,backgrounds}

\usepackage{tikz}
\usepackage{xifthen}

\pgfdeclarelayer{bg}    % declare background layer
\pgfsetlayers{bg,main}  % set the order of the layers (main is the standard layer)
\definecolor{beige}{RGB}{245, 245, 220}

\definecolor{darkgrey}{RGB}{75, 75, 75}
\definecolor{lightgrey}{RGB}{250, 250, 250}

\usetikzlibrary{calc, arrows, fit, positioning, patterns, decorations.pathreplacing, shapes}
\tikzstyle{dash} = [dashed, -latex,>=latex]
\tikzstyle{line} = [draw, -latex,>=latex]
\tikzstyle{smallbox} = [draw, minimum size=5.0mm]
\tikzstyle{box} = [draw, minimum size=7.0mm]
\tikzstyle{bigbox} = [draw, minimum size=10.0mm]
\tikzstyle{rectangle} = [draw, minimum width=10.0mm, minimum height=20.0mm]
\tikzstyle{switch} = [trapezium, trapezium angle=120, draw, rotate=90,  inner ysep=5pt, outer sep=5pt,
minimum height=7mm, minimum width=7mm]
\tikzstyle{roundbox} = [draw, circle, inner sep=0pt, minimum size=3mm]
\tikzstyle{clamped} = [draw, fill=darkgrey, minimum size=0.15cm]
\tikzstyle{msgcircle} = [shape=circle, draw, inner sep=0pt, minimum size=4mm, fill=white, font=\scriptsize]
\tikzstyle{darkmsgcircle} = [shape=circle, draw, inner sep=0pt, minimum size=4mm, fill=darkgrey, text=white, font=\scriptsize]
\tikzstyle{redmsgcircle} = [shape=circle, draw=red, inner sep=0pt, minimum size=4mm, text=red, font=\scriptsize]
\tikzstyle{reddarkmsgcircle} = [shape=circle, draw=red, inner sep=0pt, minimum size=4mm, fill=red, text=white, font=\scriptsize]
\tikzstyle{msgdoublecircle} = [shape=circle, double, double distance=1.5pt, draw, inner sep=0pt, minimum size=5mm, fill=white]
\tikzstyle{darkmsgdoublecircle} = [shape=circle, double, double distance=1.5pt, draw, inner sep=0pt, minimum size=5mm, fill=darkgrey, text=white, font=\bfseries]
\newcommand{\msg}[6]{
      \ifthenelse{\isin{#1}{left} \AND \isin{#2}{down}}{
            \coordinate (anchor) at ($({#3})!{#5}!({#4})$);
            \node[xshift=-6.0mm] at (anchor) {#6};
            \node[xshift=-1.0mm] at (anchor) {$\downarrow$};
      }{}
      \ifthenelse{\isin{#1}{right} \AND \isin{#2}{down}}{
            \coordinate (anchor) at ($({#3})!{#5}!({#4})$);
            \node[xshift=6.0mm] at (anchor) {#6};
            \node[xshift=1.0mm] at (anchor) {$\downarrow$};
      }{}

      \ifthenelse{\isin{#1}{down} \AND \isin{#2}{right}}{
            \coordinate (anchor) at ($({#3})!{#5}!({#4})$);
            \node[ yshift=-4.0mm] at (anchor) {#6};
            \node[yshift=-1.0mm] at (anchor) {$\rightarrow$};
      }{}
      \ifthenelse{\isin{#1}{up} \AND \isin{#2}{right}}{
            \coordinate (anchor) at ($({#3})!{#5}!({#4})$);
            \node[ yshift=4.0mm] at (anchor) {#6};
            \node[yshift=1.0mm] at (anchor) {$\rightarrow$};
      }{}

      \ifthenelse{\isin{#1}{down} \AND \isin{#2}{left}}{
            \coordinate (anchor) at ($({#3})!{#5}!({#4})$);
            \node[ yshift=-4.0mm] at (anchor) {#6};
            \node[yshift=-1.0mm] at (anchor) {$\leftarrow$};
      }{}
      \ifthenelse{\isin{#1}{up} \AND \isin{#2}{left}}{
            \coordinate (anchor) at ($({#3})!{#5}!({#4})$);
            \node[ yshift=4.0mm] at (anchor) {#6};
            \node[yshift=1.0mm] at (anchor) {$\leftarrow$};
      }{}

      \ifthenelse{\isin{#1}{left} \AND \isin{#2}{up}}{
            \coordinate (anchor) at ($({#3})!{#5}!({#4})$);
            \node[ xshift=-6.0mm] at (anchor) {#6};
            \node[xshift=-1.0mm] at (anchor) {$\uparrow$};
      }{}
      \ifthenelse{\isin{#1}{right} \AND \isin{#2}{up}}{
            \coordinate (anchor) at ($({#3})!{#5}!({#4})$);
            \node[ xshift=6.0mm] at (anchor) {#6};
            \node[xshift=1.0mm] at (anchor) {$\uparrow$};
      }{}
}

\newcommand{\msgcircle}[6]{
      \ifthenelse{\isin{#1}{left} \AND \isin{#2}{down}}{
            \coordinate (anchor) at ($({#3})!{#5}!({#4})$);
            \node[msgcircle,xshift=-5.0mm] at (anchor) {#6};
            \node[xshift=-1.5mm] at (anchor) {$\downarrow$};
      }{}
      \ifthenelse{\isin{#1}{right} \AND \isin{#2}{down}}{
            \coordinate (anchor) at ($({#3})!{#5}!({#4})$);
            \node[msgcircle,xshift=5.0mm] at (anchor) {#6};
            \node[xshift=1.5mm] at (anchor) {$\downarrow$};
      }{}

      \ifthenelse{\isin{#1}{down} \AND \isin{#2}{right}}{
            \coordinate (anchor) at ($({#3})!{#5}!({#4})$);
            \node[msgcircle, yshift=-5.0mm] at (anchor) {#6};
            \node[yshift=-2.0mm] at (anchor) {$\rightarrow$};
      }{}
      \ifthenelse{\isin{#1}{up} \AND \isin{#2}{right}}{
            \coordinate (anchor) at ($({#3})!{#5}!({#4})$);
            \node[msgcircle, yshift=5.0mm] at (anchor) {#6};
            \node[yshift=2.0mm] at (anchor) {$\rightarrow$};
      }{}

      \ifthenelse{\isin{#1}{down} \AND \isin{#2}{left}}{
            \coordinate (anchor) at ($({#3})!{#5}!({#4})$);
            \node[msgcircle, yshift=-5.0mm] at (anchor) {#6};
            \node[yshift=-2.0mm] at (anchor) {$\leftarrow$};
      }{}
      \ifthenelse{\isin{#1}{up} \AND \isin{#2}{left}}{
            \coordinate (anchor) at ($({#3})!{#5}!({#4})$);
            \node[msgcircle, yshift=5.0mm] at (anchor) {#6};
            \node[yshift=2.0mm] at (anchor) {$\leftarrow$};
      }{}

      \ifthenelse{\isin{#1}{left} \AND \isin{#2}{up}}{
            \coordinate (anchor) at ($({#3})!{#5}!({#4})$);
            \node[msgcircle, xshift=-5.0mm] at (anchor) {#6};
            \node[xshift=-1.5mm] at (anchor) {$\uparrow$};
      }{}
      \ifthenelse{\isin{#1}{right} \AND \isin{#2}{up}}{
            \coordinate (anchor) at ($({#3})!{#5}!({#4})$);
            \node[msgcircle, xshift=5.0mm] at (anchor) {#6};
            \node[xshift=1.5mm] at (anchor) {$\uparrow$};
      }{}
}

\newcommand{\darkmsg}[6]{
      \ifthenelse{\isin{#1}{left} \AND \isin{#2}{down}}{
            \coordinate (anchor) at ($({#3})!{#5}!({#4})$);
            \node[darkmsgcircle, xshift=-5mm] at (anchor) {#6};
            \node[xshift=-1.5mm] at (anchor) {$\downarrow$};
      }{}
      \ifthenelse{\isin{#1}{right} \AND \isin{#2}{down}}{
            \coordinate (anchor) at ($({#3})!{#5}!({#4})$);
            \node[darkmsgcircle, xshift=5mm] at (anchor) {#6};
            \node[xshift=1.5mm] at (anchor) {$\downarrow$};
      }{}

      \ifthenelse{\isin{#1}{down} \AND \isin{#2}{right}}{
            \coordinate (anchor) at ($({#3})!{#5}!({#4})$);
            \node[darkmsgcircle, yshift=-5.0mm] at (anchor) {#6};
            \node[yshift=-2.0mm] at (anchor) {$\rightarrow$};
      }{}
      \ifthenelse{\isin{#1}{up} \AND \isin{#2}{right}}{
            \coordinate (anchor) at ($({#3})!{#5}!({#4})$);
            \node[darkmsgcircle, yshift=5.0mm] at (anchor) {#6};
            \node[yshift=2.0mm] at (anchor) {$\rightarrow$};
      }{}

      \ifthenelse{\isin{#1}{down} \AND \isin{#2}{left}}{
            \coordinate (anchor) at ($({#3})!{#5}!({#4})$);
            \node[darkmsgcircle, yshift=-5.0mm] at (anchor) {#6};
            \node[yshift=-2.0mm] at (anchor) {$\leftarrow$};
      }{}
      \ifthenelse{\isin{#1}{up} \AND \isin{#2}{left}}{
            \coordinate (anchor) at ($({#3})!{#5}!({#4})$);
            \node[darkmsgcircle, yshift=5.0mm] at (anchor) {#6};
            \node[yshift=2.0mm] at (anchor) {$\leftarrow$};
      }{}

      \ifthenelse{\isin{#1}{left} \AND \isin{#2}{up}}{
            \coordinate (anchor) at ($({#3})!{#5}!({#4})$);
            \node[darkmsgcircle, xshift=-5.0mm] at (anchor) {#6};
            \node[xshift=-1.5mm] at (anchor) {$\uparrow$};
      }{}
      \ifthenelse{\isin{#1}{right} \AND \isin{#2}{up}}{
            \coordinate (anchor) at ($({#3})!{#5}!({#4})$);
            \node[darkmsgcircle, xshift=5.0mm] at (anchor) {#6};
            \node[xshift=1.5mm] at (anchor) {$\uparrow$};
      }{}
}

\newcommand{\redbackmsg}[6]{
      \ifthenelse{\isin{#1}{left} \AND \isin{#2}{down}}{
            \coordinate (anchor) at ($({#3})!{#5}!({#4})$);
            \node[reddarkmsgcircle, xshift=-5mm] at (anchor) {#6};
            \node[xshift=-1.5mm] at (anchor) {$\downarrow$};
      }{}
      \ifthenelse{\isin{#1}{right} \AND \isin{#2}{down}}{
            \coordinate (anchor) at ($({#3})!{#5}!({#4})$);
            \node[reddarkmsgcircle, xshift=5mm] at (anchor) {#6};
            \node[xshift=1.5mm] at (anchor) {$\downarrow$};
      }{}

      \ifthenelse{\isin{#1}{down} \AND \isin{#2}{right}}{
            \coordinate (anchor) at ($({#3})!{#5}!({#4})$);
            \node[reddarkmsgcircle, yshift=-5.0mm] at (anchor) {#6};
            \node[yshift=-2.0mm] at (anchor) {$\rightarrow$};
      }{}
      \ifthenelse{\isin{#1}{up} \AND \isin{#2}{right}}{
            \coordinate (anchor) at ($({#3})!{#5}!({#4})$);
            \node[reddarkmsgcircle, yshift=5.0mm] at (anchor) {#6};
            \node[yshift=2.0mm] at (anchor) {$\rightarrow$};
      }{}

      \ifthenelse{\isin{#1}{down} \AND \isin{#2}{left}}{
            \coordinate (anchor) at ($({#3})!{#5}!({#4})$);
            \node[reddarkmsgcircle, yshift=-5.0mm] at (anchor) {#6};
            \node[yshift=-2.0mm] at (anchor) {$\leftarrow$};
      }{}
      \ifthenelse{\isin{#1}{up} \AND \isin{#2}{left}}{
            \coordinate (anchor) at ($({#3})!{#5}!({#4})$);
            \node[reddarkmsgcircle, yshift=5.0mm] at (anchor) {#6};
            \node[yshift=2.0mm] at (anchor) {$\leftarrow$};
      }{}

      \ifthenelse{\isin{#1}{left} \AND \isin{#2}{up}}{
            \coordinate (anchor) at ($({#3})!{#5}!({#4})$);
            \node[reddarkmsgcircle, xshift=-5.0mm] at (anchor) {#6};
            \node[xshift=-1.5mm] at (anchor) {$\uparrow$};
      }{}
      \ifthenelse{\isin{#1}{right} \AND \isin{#2}{up}}{
            \coordinate (anchor) at ($({#3})!{#5}!({#4})$);
            \node[reddarkmsgcircle, xshift=5.0mm] at (anchor) {#6};
            \node[xshift=1.5mm] at (anchor) {$\uparrow$};
      }{}
}

\newcommand{\redmsg}[6]{
      \ifthenelse{\isin{#1}{left} \AND \isin{#2}{down}}{
            \coordinate (anchor) at ($({#3})!{#5}!({#4})$);
            \node[redmsgcircle, xshift=-5mm] at (anchor) {#6};
            \node[xshift=-1.5mm] at (anchor) {$\downarrow$};
      }{}
      \ifthenelse{\isin{#1}{right} \AND \isin{#2}{down}}{
            \coordinate (anchor) at ($({#3})!{#5}!({#4})$);
            \node[redmsgcircle, xshift=5mm] at (anchor) {#6};
            \node[xshift=1.5mm] at (anchor) {$\downarrow$};
      }{}

      \ifthenelse{\isin{#1}{down} \AND \isin{#2}{right}}{
            \coordinate (anchor) at ($({#3})!{#5}!({#4})$);
            \node[redmsgcircle, yshift=-5.0mm] at (anchor) {#6};
            \node[yshift=-2.0mm] at (anchor) {$\rightarrow$};
      }{}
      \ifthenelse{\isin{#1}{up} \AND \isin{#2}{right}}{
            \coordinate (anchor) at ($({#3})!{#5}!({#4})$);
            \node[redmsgcircle, yshift=5.0mm] at (anchor) {#6};
            \node[yshift=2.0mm] at (anchor) {$\rightarrow$};
      }{}

      \ifthenelse{\isin{#1}{down} \AND \isin{#2}{left}}{
            \coordinate (anchor) at ($({#3})!{#5}!({#4})$);
            \node[redmsgcircle, yshift=-5.0mm] at (anchor) {#6};
            \node[yshift=-2.0mm] at (anchor) {$\leftarrow$};
      }{}
      \ifthenelse{\isin{#1}{up} \AND \isin{#2}{left}}{
            \coordinate (anchor) at ($({#3})!{#5}!({#4})$);
            \node[redmsgcircle, yshift=5.0mm] at (anchor) {#6};
            \node[yshift=2.0mm] at (anchor) {$\leftarrow$};
      }{}

      \ifthenelse{\isin{#1}{left} \AND \isin{#2}{up}}{
            \coordinate (anchor) at ($({#3})!{#5}!({#4})$);
            \node[redmsgcircle, xshift=-5.0mm] at (anchor) {#6};
            \node[xshift=-1.5mm] at (anchor) {$\uparrow$};
      }{}
      \ifthenelse{\isin{#1}{right} \AND \isin{#2}{up}}{
            \coordinate (anchor) at ($({#3})!{#5}!({#4})$);
            \node[redmsgcircle, xshift=5.0mm] at (anchor) {#6};
            \node[xshift=1.5mm] at (anchor) {$\uparrow$};
      }{}
}

\newcommand{\bwmsg}[6]{
      \ifthenelse{\isin{#1}{left} \AND \isin{#2}{down}}{
            \coordinate (anchor) at ($({#3})!{#5}!({#4})$);
            \node[msgdoublecircle, xshift=-5.5mm] at (anchor) {#6};
            \node[xshift=-1.5mm] at (anchor) {$\downarrow$};
      }{}
      \ifthenelse{\isin{#1}{right} \AND \isin{#2}{down}}{
            \coordinate (anchor) at ($({#3})!{#5}!({#4})$);
            \node[msgdoublecircle, xshift=5.5mm] at (anchor) {#6};
            \node[xshift=1.5mm] at (anchor) {$\downarrow$};
      }{}

      \ifthenelse{\isin{#1}{down} \AND \isin{#2}{right}}{
            \coordinate (anchor) at ($({#3})!{#5}!({#4})$);
            \node[msgdoublecircle, yshift=-6.0mm] at (anchor) {#6};
            \node[yshift=-2.0mm] at (anchor) {$\rightarrow$};
      }{}
      \ifthenelse{\isin{#1}{up} \AND \isin{#2}{right}}{
            \coordinate (anchor) at ($({#3})!{#5}!({#4})$);
            \node[msgdoublecircle, yshift=6.0mm] at (anchor) {#6};
            \node[yshift=2.0mm] at (anchor) {$\rightarrow$};
      }{}

      \ifthenelse{\isin{#1}{down} \AND \isin{#2}{left}}{
            \coordinate (anchor) at ($({#3})!{#5}!({#4})$);
            \node[msgdoublecircle, yshift=-6.0mm] at (anchor) {#6};
            \node[yshift=-2.0mm] at (anchor) {$\leftarrow$};
      }{}
      \ifthenelse{\isin{#1}{up} \AND \isin{#2}{left}}{
            \coordinate (anchor) at ($({#3})!{#5}!({#4})$);
            \node[msgdoublecircle, yshift=6.0mm] at (anchor) {#6};
            \node[yshift=2.0mm] at (anchor) {$\leftarrow$};
      }{}

      \ifthenelse{\isin{#1}{left} \AND \isin{#2}{up}}{
            \coordinate (anchor) at ($({#3})!{#5}!({#4})$);
            \node[msgdoublecircle, xshift=-5.5mm] at (anchor) {#6};
            \node[xshift=-1.5mm] at (anchor) {$\uparrow$};
      }{}
      \ifthenelse{\isin{#1}{right} \AND \isin{#2}{up}}{
            \coordinate (anchor) at ($({#3})!{#5}!({#4})$);
            \node[msgdoublecircle, xshift=5.5mm] at (anchor) {#6};
            \node[xshift=1.5mm] at (anchor) {$\uparrow$};
      }{}
}

\newcommand{\bwdarkmsg}[6]{
      \ifthenelse{\isin{#1}{left} \AND \isin{#2}{down}}{
            \coordinate (anchor) at ($({#3})!{#5}!({#4})$);
            \node[darkmsgdoublecircle, xshift=-5.5mm] at (anchor) {#6};
            \node[xshift=-1.5mm] at (anchor) {$\downarrow$};
      }{}
      \ifthenelse{\isin{#1}{right} \AND \isin{#2}{down}}{
            \coordinate (anchor) at ($({#3})!{#5}!({#4})$);
            \node[darkmsgdoublecircle, xshift=5.5mm] at (anchor) {#6};
            \node[xshift=1.5mm] at (anchor) {$\downarrow$};
      }{}

      \ifthenelse{\isin{#1}{down} \AND \isin{#2}{right}}{
            \coordinate (anchor) at ($({#3})!{#5}!({#4})$);
            \node[darkmsgdoublecircle, yshift=-6.0mm] at (anchor) {#6};
            \node[yshift=-2.0mm] at (anchor) {$\rightarrow$};
      }{}
      \ifthenelse{\isin{#1}{up} \AND \isin{#2}{right}}{
            \coordinate (anchor) at ($({#3})!{#5}!({#4})$);
            \node[darkmsgdoublecircle, yshift=6.0mm] at (anchor) {#6};
            \node[yshift=2.0mm] at (anchor) {$\rightarrow$};
      }{}

      \ifthenelse{\isin{#1}{down} \AND \isin{#2}{left}}{
            \coordinate (anchor) at ($({#3})!{#5}!({#4})$);
            \node[darkmsgdoublecircle, yshift=-6.0mm] at (anchor) {#6};
            \node[yshift=-2.0mm] at (anchor) {$\leftarrow$};
      }{}
      \ifthenelse{\isin{#1}{up} \AND \isin{#2}{left}}{
            \coordinate (anchor) at ($({#3})!{#5}!({#4})$);
            \node[darkmsgdoublecircle, yshift=6.0mm] at (anchor) {#6};
            \node[yshift=2.0mm] at (anchor) {$\leftarrow$};
      }{}

      \ifthenelse{\isin{#1}{left} \AND \isin{#2}{up}}{
            \coordinate (anchor) at ($({#3})!{#5}!({#4})$);
            \node[darkmsgdoublecircle, xshift=-5.5mm] at (anchor) {#6};
            \node[xshift=-1.5mm] at (anchor) {$\uparrow$};
      }{}
      \ifthenelse{\isin{#1}{right} \AND \isin{#2}{up}}{
            \coordinate (anchor) at ($({#3})!{#5}!({#4})$);
            \node[darkmsgdoublecircle, xshift=5.5mm] at (anchor) {#6};
            \node[xshift=1.5mm] at (anchor) {$\uparrow$};
      }{}
}

\tikzset{mainstyle/.style={fill=white, draw=black, shape=rectangle, align=center}}

\tikzset{dstyle/.style={mainstyle, minimum size=4mm, inner sep=0pt, text width=4mm}}

\tikzset{sstyle/.style={mainstyle, minimum size=5mm, inner sep=0pt, text width=5mm}}

\tikzset{ostyle/.style={fill=darkgrey, draw=black, shape=rectangle, minimum size=0.2cm, inner sep=0pt, text width=2mm}}

\tikzstyle{observation}=[ostyle]
\tikzstyle{deterministic}=[dstyle]
\tikzstyle{stochastic}=[sstyle]

\tikzstyle{filter}=[mainstyle, minimum width=1cm, minimum height=0.5cm]
\tikzstyle{selector}=[fill=white, draw=black, shape=trapezium, rotate=180, minimum width=1cm, minimum height=0.5cm]

\DeclareRobustCommand{\cev}[1]{%
  \mathpalette\do@cev{#1}%
}
\newcommand{\do@cev}[2]{%
  \fix@cev{#1}{+}%
  \reflectbox{$\m@th#1\vec{\reflectbox{$\fix@cev{#1}{-}\m@th#1#2\fix@cev{#1}{+}$}}$}%
  \fix@cev{#1}{-}%
}
\newcommand{\fix@cev}[2]{%
  \ifx#1\displaystyle
    \mkern#23mu
  \else
    \ifx#1\textstyle
      \mkern#23mu
    \else
      \ifx#1\scriptstyle
        \mkern#22mu
      \else
        \mkern#22mu
      \fi
    \fi
  \fi
}

\usepackage[disable]{todonotes}
\newcommand{\bdv}[2][] {\todo[inline,backgroundcolor=blue!20!white, #1]{(Bert) #2}}
\newcommand{\sepideh}[2][] {\todo[inline,backgroundcolor=green!20!white, #1]{(Sepideh) #2}}

\begin{document}
\title{Spike-Timing-Dependent Plasticity \\ for Bernoulli Message Passing}
\author{Sepideh Adamiat\inst{1} \and
Wouter M. Kouw\inst{1} \and
Bert de Vries\inst{1,2}
}
\authorrunning{Adamiat et al.}
\institute{Electrical Engineering Department, TU Eindhoven, Netherlands \and
Lazy Dynamics B.V., Eindhoven, Netherlands \\
\email{s.adamiat@tue.nl}
}
% \author{Anonymous Authors}
%
\maketitle              % typeset the header of the contribution
\begin{abstract}
Bayesian inference provides a principled framework for understanding brain function, while neural activity in the brain is inherently spike-based. This paper bridges these two perspectives by designing spiking neural networks that simulate Bayesian inference through message passing for Bernoulli messages. To train the networks, we employ spike-timing-dependent plasticity, a biologically plausible mechanism for synaptic plasticity which is based on the Hebbian rule. Our results demonstrate that the network's performance closely matches the true numerical solution. We further demonstrate the versatility of our approach by implementing a factor graph example from coding theory, illustrating signal transmission over an unreliable channel.

\end{abstract}

\section{Introduction}

Numerous perceptual and motor tasks carried out by the human nervous system can be effectively described using a Bayesian inference framework \cite{knill1996perception,fiorillo2003discrete}. According to the Bayesian brain hypothesis, the brain updates its beliefs about the world by integrating sensory input with prior knowledge~\cite{doya2007bayesian}. This concept aligns with the Free Energy Principle (FEP), which asserts that living systems adapt to their environment by maximizing evidence for an internal generative model of sensory observations \cite{friston2006free}. This adaptation process is conceptually carried out by variational free energy minimization, physically realized by exchanging action potentials between neurons.  

Recent studies have explored the information processing capacity of in vitro biological neurons cultured on top of multi-electrode arrays, demonstrating their potential for unsupervised learning, speech recognition, and decision-making tasks \cite{cai2023brain}. Some of these studies have employed the Free Energy Principle, including research by \cite{kagan2022vitro}, which harnesses the inherent adaptive computation of neurons in a simulated game world, and \cite{isomura2015cultured}, which investigates their application in blind source separation. These studies are evidence supporting the FEP and Bayesian Brain hypothesis.

Bayesian inference can be computationally intensive due to the complexity of integrating and marginalizing over probability distributions. Graphical representations help to manage this complexity by breaking the problem into smaller, localized computations. These graphical models provide a structured framework for inference, where message-passing algorithms (also known as belief propagation) efficiently compute posterior probabilities by passing information along the graph's edges. This approach simplifies inference for large and sparsely connected systems, making Bayesian methods suitable for practical applications \cite{loeliger_factor_2007}.
% Interestingly, the FEP can also be formulated as message passing under a generative model \cite{friston2017graphical}. 
Interestingly, active inference, which is a corollary of FEP, can also be realized by message passing on a factor graph representation of the generative model \cite{friston2017graphical}.

\bdv{Let's be precise. The FEP is a principle and cannot be formulated as MP. Active inference is a process that is consistent with FEP, and (being an inference process) it can be realized by message passing on a factor graph representation of the generative model.}\sepideh{Changed.}
On the other hand, from a biological point of view, all electrical communication in the brain is realized by spike-based message passing, and to understand the brain's behavior or communicate with cultured neurons, we need to encode and decode information in spike forms. Spiking Neural Networks (SNNs) are a class of artificial neural networks that provide a more accurate simulation of neural processes, making them significant for advances in computational neuroscience \cite{maass1997networks}. These biologically inspired neural models communicate with each other by discrete spike trains as input and output. There are several advances in using SNNs in machine learning tasks, including \cite{wu2022training}, which implemented reinforcement learning to control an inverted pendulum problem, and \cite{lobov2020spatial} that proposed a self-learning spiking network to control a mobile robot. These studies highlight the potential of SNNs for robotic control and autonomous decision making. SNNs are also recognized for their efficient energy consumption when running on neuromorphic devices \cite{yamazaki2022spiking}.

Numerous studies have explored the biological plausibility of message passing algorithms \cite{parr2019neuronal}. In \cite{maass2011liquid}, an innovative approach to implementing sum-product message passing within spiking neural networks is introduced. This approach utilizes a network of interconnected liquid state machines. While their work offers valuable insights into bridging the gap between SNNs and message passing, it does not address the implementation of a learning mechanism for synaptic weights—a gap that this paper aims to fill.

This paper presents a method for training  SNNs with spike-timing-dependent plasticity (STDP), a biologically inspired synaptic update rule widely used in neuromorphic computing, to implement sum-product message passing for Bernoulli-distributed messages. We demonstrate that the proposed networks, constructed with a minimal number of neurons and synaptic connections, yield results closely aligned with numerical results. Furthermore, we apply the proposed network to an example of a noisy signal transmission channel, demonstrating the principles and generality of our approach.
\section{Background}\label{sec:background}

In this section, we outline two key concepts we aim to connect: Bayesian inference with message passing and spiking neural networks. Readers already familiar with these topics may choose to skip this section.

\subsection{Message Passing on Forney-style Factor Graphs}
\begin{figure}[tb]
\centerline{\begin{tikzpicture}

\node[smallbox] (g) {$=$};
\node[smallbox,left = 10 mm  of g] (f_b) {$f_b$};
\node[smallbox,below = 7 mm  of g] (f_c) {$f_c$};
\node[smallbox,right = 10 mm  of g] (f_d) {$f_d$};
\node[smallbox,left = 18 mm  of f_b] (x_1) {$f_a$};
\node[smallbox,left = 10 mm  of f_c] (x_3) {$f_e$};
\node[smallbox,below = 10 mm  of f_c] (x_4) {$f_f$};

\draw[->] (f_b) -- (g)node[pos=0.5, above] {$x_2$};
\draw[->] (x_1) -- (f_b)node[pos=0.5, above] {$x_1$}node[pos=0.75, below] {\tiny{$\overrightarrow{\mu}(x_1)$}}node[pos=0.25, below] {\tiny$\overleftarrow{\mu}(x_1)$};
\draw[<-] (g) -- (f_c)node[pos=0.5, left] {$x_2'$};
\draw[->] (x_3) -- (f_c)node[pos=0.5, above] {$x_3$};
\draw[->] (x_4) -- (f_c)node[pos=0.5, left] {$x_4$} node[pos=0.7, right] {\tiny{$\uparrow \overrightarrow{\mu}(x_4)$}}node[pos=0.3, right] {\tiny$\downarrow\overleftarrow{\mu}(x_4)$};
\draw[->] (g) -- (f_d)node[pos=0.5, above] {$x_2''$};

\end{tikzpicture}}
\caption{The FFG corresponding to the example model defined in Equation \eqref{eq:Model-example1}.
\bdv{If you write the direction of a message with an overhead arrow, then the edges on the graph must be directed! Check this also for some other figures where this is also wrong. If you use undirected edges, you would need to write the direction as $\mu_{ f_f\rightarrow f_c}(x_4)$, which (in my opinion) is too verbose, so I recommend adding directions to the edges. Your proposal below (forward message corresponds to from-left-to-right) is not consistent with the literature and fragile.}\sepideh{Done.} }
\label{fig: mpexample}
\end{figure}
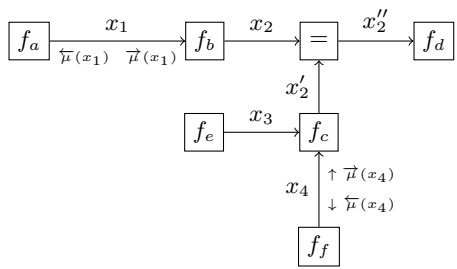

 Sum-product message passing, or belief propagation, in Forney-style Factor Graphs (FFGs) is a powerful algorithm for performing Bayesian inference. A large variety of algorithms in fields of machine learning, signal processing, coding, and statistics may be viewed as special cases of this method \cite{loeliger_factor_2007} \cite{loeliger2004introduction} \cite{palmieri2022unifying}.
An FFG offers a graphical description of a factorized function \cite{forney2001codes}.
In an FFG, nodes represent functions, and edges represent variables. An edge connects to a node if and only if it represents a variable that is an argument of the node's function. As an example, consider the factorized function
\begin{align}\label{eq:Model-example1}
     f(x_1, x_2,x_3,x_4)  = f_a(x_1) f_b(x_1,x_2)f_c(x_2, x_3,x_4)f_d(x_2)f_e(x_3)f_f(x_4),
\end{align}
with its corresponding FFG is illustrated in Figure~\ref{fig: mpexample}. Note that in an FFG, each edge can maximally connect to two nodes. 
If a variable is an argument in more than two factors, we use an "equality" node $f(x,y,z) = \delta(z-y)\delta(z-x)$ that enforces the same beliefs across all variables ($x$, $y$ and $z$) that connect to the equality node. 

The sum-product algorithm passes "messages", which are probabilistic distributions, along the graph's edges. This method is an efficient approach for conducting probabilistic inference within sparsely connected generative models. Message-passing-based inference is particularly scalable because it exploits the model's independence structure, significantly reducing the computational complexity compared to naive Bayesian inference, which requires summing over all possible configurations globally. The local and distributed nature of the sum-product algorithm allows it to handle large-scale problems efficiently, making it widely applicable in areas such as multi-agent trajectory planning~\cite{VanErp:inpress}, control of non-linear systems \cite{adamiat2024message} and active inference in non-linear environments~\cite{van2019simulating}. 

% Messages pass in both directions along the edges of the FFG, we denote messages going left to right or down to up with the notation $\overrightarrow{\mu}(\cdot)$, while messages going in the opposite directions are represented by $\overleftarrow{\mu}(\cdot)$. 
In FFG, messages are passed bidirectionally along the edges. For notational convenience, we assign a direction to each edge and denote messages propagating in the designated direction by $\overrightarrow{\mu}(\cdot)$, while those traveling in the opposite direction are represented by $\overleftarrow{\mu}(\cdot)$.
\bdv{Please stick to factor graph notational conventions (use directed edges and overhead arrows on the messages) rather than inventing your own. In your proposal, if you rotate the paper, forward becomes backward. Not a good convention. I have made a comment on that in a previous version.}\sepideh{Done.}
In general, for any node $f(y,x_1,\dots,x_n)$, 
the sum-product rule for an outgoing message over edge $y$ is given by

\begin{equation}\label{eq:sum-product}
    \underbrace{\overrightarrow\mu_y({y})}_{\text{outgoing messages}} =  \int \underbrace{\overrightarrow\mu_{x_1}({x}_1)...\overrightarrow\mu_{x_n}({x}_n) }_{\text{incoming messages}}\underbrace{f({y},{x}_1,...,{x}_n)}_{\text{node function}} \mathrm{d}{x}_1...\mathrm{d}{x}_n .
\end{equation}

 By pre-calculating the update rules for common model components, one can efficiently construct and adjust inference algorithms without extensive computation. Certain studies have proposed message update rules \cite{loeliger_factor_2007,winn2005variational}, while there exist software toolboxes that provide the pre-computed message update rules for commonly used distributions and factors \cite{bagaev2023reactive,InferNET18} to automate the inference process for making applications. Tables \ref{table:mp-b} present some key update rules for processing incoming messages that carry a Bernoulli distribution.
 \bdv{be precise: ... present some key update rules for processing incoming messages that carry a Bernoulli distribution.}
 \sepideh{Done.}
 In this study, we aim to implement these computations using spiking neural networks, offering a biologically inspired approach to probabilistic inference.

 \begin{table}[htbp]
\caption{Sum-product message update rules for the equality and XOR nodes, for given input messages $\overrightarrow\mu_x(x) = \mathcal{B}er(x | p_x)$ and $\overrightarrow\mu_y(y) = \mathcal{B}er(y | p_y)$\cite{loeliger2004introduction}. \bdv{figure is not complete: the RHS defines a message (eg $\overrightarrow\mu_z(z)$, and the corresponding graph at LHS shows no direction of forward vs backward, nor are there any messages to be seen on the graph. }\sepideh{Directions are added. The notation of messages matches the literature \cite{loeliger2004introduction}.} \bdv{I still don't see the messages $\overrightarrow\mu_x(x)$, $\overrightarrow\mu_y(y)$ and $\overrightarrow\mu_z(z)$ in the pictures.}
}
\begin{center}
\setlength{\tabcolsep}{15pt}
\begin{tabular}{|c|c|}
\hline
% \multicolumn{3}{|c|}{\textbf{An unreliable channel model}}\\
% \cline{1-3} 
 \textbf{\textit{Node function}}& \textbf{\textit{Update rule}} \\
\hline
  \begin{tabular}{c}\\ \begin{tikzpicture}
\node[smallbox] (g) {$=$};
\node[below = 4 mm  of g] (down) {};
\node[left = 5 mm  of g] (left) {};
\node[right = 5 mm  of g] (right) {};

\draw[->] (left) -- (g)node[pos=0.5, above] {$X$};
\draw[->] (g) -- (right)node[pos=0.5, above] {$Z$};;
\draw[<-] (g) -- (down)node[pos=0.5, right] {$Y$};;
\end{tikzpicture} \end{tabular}& \begin{tabular}{c} 
  \normalsize $\overrightarrow\mu_z(z) = \mathcal{B}er(z \, | \, \frac{p_xp_y}{1 -p_x+2p_x p_y-p_y})$\\  \\  \end{tabular} \\
  \hline
  \begin{tabular}{c}\\ \begin{tikzpicture}
\node[smallbox] (g) {XOR};
\node[below = 5 mm  of g] (down) {};
\node[left = 5 mm  of g] (left) {};
\node[right = 5 mm  of g] (right) {};

\draw[->] (left) -- (g)node[pos=0.5, above] {$X$};
\draw[->] (g) -- (right)node[pos=0.5, above] {$Z$};;
\draw[<-] (g) -- (down)node[pos=0.5, right] {$Y$};;
\end{tikzpicture}\end{tabular} &  \begin{tabular}{c} \normalsize $ \overrightarrow\mu_z(z) = \mathcal{B}er(z \, | \, p_x-2p_x p_y+p_y)\,$\\ \\  \end{tabular}\\
\hline
\end{tabular}
\label{table:mp-b}
\end{center}
\end{table}

\subsection{Spiking Neural Networks}

SNNs are a class of artificial neural networks that more closely mimic biological neural systems compared to traditional artificial neural networks. Unlike standard models that use continuous activation values, SNNs process and transmit information using discrete spike events over time. This event-driven paradigm makes SNNs well-suited for combining with MP, and also enables more energy-efficient computation.

\subsubsection{Neuron Models}

SNNs employ neuron models that describe the dynamics of membrane potential and spike generation. One of the most commonly used models is the Leaky Integrate-and-Fire (LIF) neuron, which approximates the behavior of a biological neuron with a differential equation. The membrane potential \( V(t) \) evolves according to

\[
\tau_m \frac{dV(t)}{dt} = - (V(t) - V_{\text{rest}}) + R I(t) \,,
\]
where \( \tau_m \) is the membrane time constant,
 \( V_{\text{rest}} \) is the resting membrane potential, \( R \) is the membrane resistance, \( I(t) \) is the synaptic input current \cite{gerstner2002spiking}. \( I(t) \) represents the weighted sum of spikes from presynaptic neurons, capturing the total synaptic input to the neuron.

The leak mechanism refers to the gradual decay of the membrane potential \( V(t) \) back toward its resting value \( V_{\text{rest}} \) in the absence of input. This models the natural tendency of neurons to return to a stable baseline and prevents indefinite accumulation of input. 
The integrating aspect represents the accumulation of incoming currents \( I(t) \), which drive \( V(t) \) upward. \bdv{Generally speaking, current is the result of a voltage difference. You make it seem here as if the current is some autonomous causal process.} When the membrane potential reaches a certain threshold \( V_{\text{th}} \), the neuron fires: it emits a spike (action potential), and the membrane potential is immediately reset to a lower value \( V_{\text{reset}} \). This is the fire mechanism, mimicking the all-or-nothing nature of biological spikes.
After firing, the neuron may enter a refractory period during which it is temporarily unable to spike, ensuring separation between consecutive spikes and limiting firing rates.

\bdv{So, part of the spike generating process is represented by an (unnumbered) equation for membrane potentials, and part is just hidden in some textual description that introduces new variables on the fly. Probably too late to repair for IWAI submission, but in a future version of this paper you should present a proper model for spike generation, not some equation for membrane potentials, and then some text. There are actual proper models. In particular, \url{https://arxiv.org/abs/2410.19315} might be relevant in this context.}

\subsubsection{Synaptic Models}\label{stdp}

Synaptic models in SNNs determine how spikes from presynaptic neurons affect the membrane potential of postsynaptic neurons. In addition to shaping input currents, synapses can also adapt over time through learning mechanisms. One biologically inspired form of synaptic plasticity is STDP, which adjusts synaptic weights based on the relative timing of pre- and postsynaptic spikes.The change in synaptic weight \( \Delta w \) is typically modeled as

\[
\Delta w =
\begin{cases}
A_+ \exp\left(-\frac{\Delta t}{\tau_+}\right) & \text{if } \Delta t \geq 0 \\
- A_- \exp\left(\frac{\Delta t}{\tau_-}\right) & \text{if } \Delta t < 0 ,
\end{cases}
\]
\bdv{equations should be inserted in a sentence as if they are nouns. So, you should not use a semicolon here. You should really know this by now and take care that your papers do not contain those types of errors anymore.}\sepideh{Done.}
\bdv{What happened to the case $\Delta t = 0$? I remember I have made a comment on that in a previous version. }\sepideh{In some references, including the one cited in this paper, the case $\Delta t = 0$ is not explicitly defined. However, other works include it under the condition $\Delta t \geq 0$. In practice, exact coincidence of pre- and postsynaptic spikes$\Delta t = 0$ is rare and often not considered. To avoid confusion, I adopt the convention $\Delta t \geq 0$ in this work.}
\bdv{As a general point of feedback, if I make a comment in a paper, I prefer that you either accept it or let me know why you disagree (which is OK, this is \emph{your} paper), but don't delete comments without any answer, because then I have to do the same reviews over and again.}\sepideh{Certainly. I never delete any of your comments. You made similar remarks on another paper of mine, which is related to this one but focuses on Gaussian messages. I have addressed those comments there—or I am in the process of doing so. This overlap may be the source of the confusion.} where \( \Delta t = t_{\text{post}} - t_{\text{pre}} \) is the time difference between the postsynaptic and presynaptic spikes, \( A_+ \) and \( A_- \) are learning rates, and \( \tau_+ \) and \( \tau_- \) are time constants for potentiation and depression, respectively \cite{song2000competitive}.

In summary, STDP strengthens synapses when the presynaptic neuron fires shortly before the postsynaptic neuron and weakens them when the order is reversed, thus encoding temporal correlations in spike activity.

\section{Methodology}\label{sec: method}

In this section, we introduce networks of LIF neurons trained with the biologically plausible synaptic learning rule, STDP, to simulate factor nodes in the message passing algorithm. This is achieved by encoding Bernoulli distributions into spike trains, passing them through the proposed networks, and decoding the output spike trains back into Bernoulli distributions. The output messages are compared with the numerical results derived from the sum-product rule~\eqref{eq:sum-product}. The code used to produce the results in this paper is available at \url{https://github.com/biaslab/stdp-bernoulli-message-passing}.

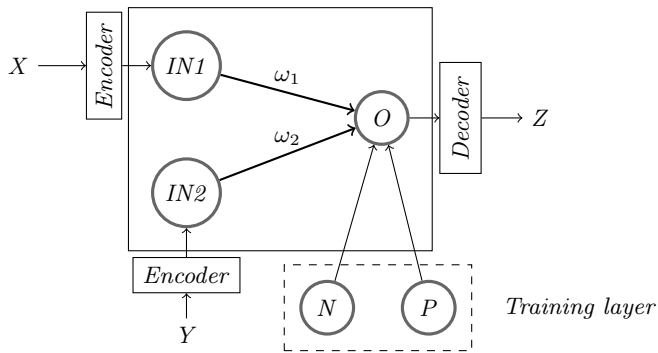
\begin{figure*}[tb]
\centerline{\begin{tikzpicture}[
roundnode/.style={circle, draw=black!60, very thick, minimum size=7mm},
]
% logical Node
\node[roundnode] (Out) {\textit{O}};

\node[roundnode, above left = 1mm and 20 mm  of Out] (up) {\textit{IN1}};
\node[roundnode,below left = 4 mm and 20 mm of Out] (down) {\textit{IN2}};

\node[left = 15 mm  of up] (x) {\textit{X}};
\node[left = 4 mm  of up, rotate=90, anchor=south] (en_x) {\textit{Encoder}};
\node[fit=(en_x), draw, inner sep = .1 mm] (box_en_x) {};

\node[below = 12 mm  of down] (y) {\textit{Y}};
\node[below = 4 mm  of down] (en_y) {\textit{Encoder}};
\node[fit=(en_y), draw, inner sep = .1 mm] (box_en_y) {};

\node[right = 15 mm  of Out] (z) {\textit{Z}};
\node[right = 9 mm  of Out, rotate=90, anchor=south] (dec) {\textit{Decoder}};
\node[fit=(dec), draw, inner sep = .5 mm] (box_d) {};

\node[roundnode,  below right = 20 mm and 1 mm  of Out] (p) {\textit{P}};
\node[roundnode,  below left = 20 mm and 2 mm of Out] (n) {\textit{N}};

% \node[above = 15 mm  of Out] (title2) {\textit{Output layer}};
% \node[left = 5 mm  of title2] (title1) {\textit{Input layer}};
\node[right = 5 mm  of p] (title3) {\textit{Training layer}};

\node[fit=(p)(n), draw, inner sep = 2 mm, dashed] (box) {};
\node[fit=(up)(down)(Out), draw, inner sep = 3 mm] (box) {};

\draw[->, thick] (up) -- (Out)node[pos=0.5, above] {$\omega_{1} $};
\draw[->, thick] (down) -- (Out) node[pos=0.5, above] {$\omega_{2}$};

\draw[->] (p) -- (Out)node[pos=0.7, above] {};
\draw[->] (n) -- (Out)node[pos=0.5, right] {};

\draw[->] (x) -- (box_en_x);
\draw[->] (box_en_x) -- (up);

\draw[->] (y) -- (box_en_y);
\draw[->] (box_en_y) -- (down);

\draw[->] (box_d) -- (z);
\draw[->] (Out) -- (box_d);

\end{tikzpicture}}
\caption{SNN architecture for simulating the sum-product update rules of AND, OR, and NOT factor nodes, as represented in Table~\ref{table:mp-logic}. Random variables $X$ and $Y$ are encoded into spike trains, and the resulting output spike train is decoded into $Z$. Each circle represents a LIF neuron. The training layer is removed after training the synaptic weights $\omega_1$ and $\omega_2$ using the STDP algorithm. For computing NOT of $X$, the variable $Y$ is configured to emit constant spikes at every time step by defining its Bernoulli distribution as $\overrightarrow{\mu}_y(y) = \mathcal{B}er(y \mid p_y = 1)$. \bdv{How should I read this graph? Is this a bayesian network? Why is $Y$ coming from the bottom and $X$ from the left? Is there an asymmetry for $X$ and $Y$?}\sepideh{The notation is used in this way to match the nodes defined in table 4.} \bdv{What is "constant spikes"?}\sepideh{I modified the text a little.}}
\label{fig: logical_gates1}
\end{figure*}

Since this study focuses on Bernoulli messages, we start with basic logical operations AND, OR, and NOT, shown in Table \ref{table:truth}. Functionally complete sets, such as \{AND, NOT\} or \{OR, NOT\}, serve as the foundation of logical computation, allowing for the construction of any logical operation through combinations of these primitives \cite{enderton2001mathematical}. As a further example, we implement the XOR factor node using a combination of the AND, OR, and NOT networks.
\begin{table}[htbp]
\vspace{-10pt}
\caption{Logical Gates. The logical relationships defined here specify the desired output signals for training the networks using STDP. \bdv{This caption is insufficient. What is the purpose of this table?}\sepideh{Done.}}
\begin{center}
\setlength{\tabcolsep}{15pt}
\begin{tabular}{|c|c|c|c|c|c|c|}
\hline
 \textbf{\textit{$S_1\quad \, S_2$}}& \textbf{\textit{OR}} & \textbf{\textit{AND }}& \textbf{\textit{NOT-$S_1$}}& \textbf{\textit{XOR}}\\
\hline
 $0\quad  \quad0$   & 0   & 0 &  1 & 0\\

 $0\quad  \quad1$   & 1   & 0 &  1 & 1\\
 
 $1\quad  \quad0$   & 1   & 0 &  0 & 1\\
 
 $1\quad  \quad1$   & 1   & 1 &  0 & 0\\
\hline
\end{tabular}
\label{table:truth}
\end{center}
\vspace{-10pt}
\end{table}
The architecture of the proposed spiking neural network is illustrated in Figure \ref{fig: logical_gates1}. Each circle represents a LIF neuron. The network consists of three layers: input, output, and a temporary training layer. The input layer comprises two neurons, each receiving encoded spike trains of the input Bernoulli messages. To encode Bernoulli distributions as spike trains, we sample them at a rate of $100$ samples per second (i.e., every $10$ ms). These input neurons are connected to the output neuron via trainable synaptic weights, denoted as $\omega_1$ and $\omega_2$, which are trained to produce the desired logical output.

To facilitate training, a training layer is used to activate the output neuron during learning; this layer is removed after training is complete. According to the STDP learning rule, synapses are strengthened when the output neuron fires shortly after the input spikes. During training, we generate the correct output spike trains based on the truth table shown in Table \ref{table:truth}.
These target spike trains are delayed by $1$ ms and provided to the network via the positive training neuron, to activate the output neuron slightly after the input neurons. To weaken synapses when no spike is expected in the true output, we preemptively activate the output neuron $1$ ms before the input spikes occur. This mechanism ensures that STDP decreases the synaptic strength in undesired scenarios.

\begin{table}[ht]
\centering
\caption{The parameter values used for LIF and STDP in section \ref{sec: method}. \bdv{refer to the experimental section where the experiment is described. More generally, this table is hard to interpret. What should we think of $\text{lr * w\textsubscript{max} = 0.005}$?}\sepideh{Done.}}
\setlength{\tabcolsep}{15pt}
\begin{tabular}{|c|c|c|c|c|c|c|}
\hline
\textbf{V\textsubscript{th}} & \textbf{V\textsubscript{rest}} & \textbf{$\tau$\textsubscript{m}} & \textbf{R} & \textbf{w\textsubscript{max}} & \textbf{w\textsubscript{min}} \\ \hline
\text{-50 mV}                 & \text{-80 mV}                  & \text{5 ms}    &       \text{1}        & \text{1}                      & \text{-1}   \\ \hline
\textbf{a\textsuperscript{+}} & \textbf{a\textsuperscript{-}}   & \textbf{$\tau$\textsuperscript{+}}   & \textbf{$\tau$\textsuperscript{-}}  & \textbf{t\textsubscript{interval}} & \textbf{lr}  \\ \hline
\text{0.005} & \text{-a\textsuperscript{+}} & \text{20 ms} & \text{20 ms} & \text{5 ms} &  \text{0.005}\\ \hline
\end{tabular}
\label{table:parameters}
\end{table}

\begin{figure}[tb]
    \includegraphics[width=1\columnwidth]{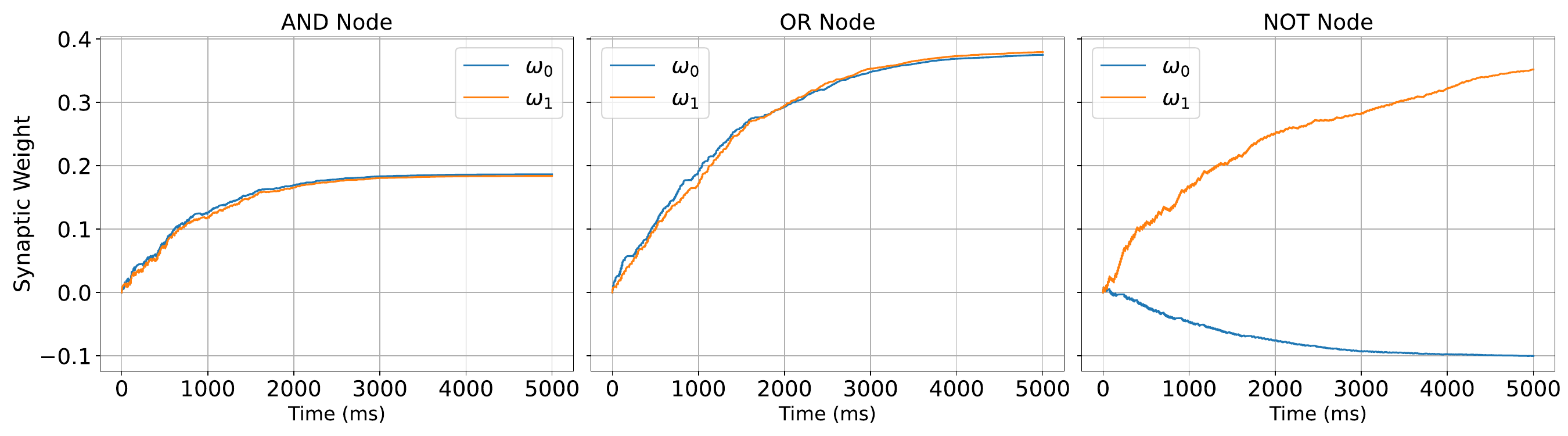}
  \caption{Evolution of synaptic weights for AND, OR, and NOT nodes, from STDP-based training described in Section~\ref{sec: method}.\bdv{In what experiment? Refer to the textual passage where you present this.}}
  \label{fig:STDP_logic_nodes}
\end{figure}

The evolution of the synaptic weights $\omega_1$ and $\omega_2$ during training is shown in Figure \ref{fig:STDP_logic_nodes}. This training approach is inspired by \cite{mo2021logicsnn}, and all neuron and synapse model parameters follow that study. The parameter values are summarized in Table \ref{table:parameters}. We used the \textrm{Brian2} toolbox ~\cite{goodman2008brian} to simulate neural activities.

For training, the input probabilities $p_x$ and $p_2$ are initially set to 0.5. However, for the NOT operation, where the goal is to compute the negation of input 1, we fix $p_y = 1$, ensuring that the second input neuron fires at every time step. As shown in the figure, the network learns to strengthen the synaptic connection $\omega_2$ while weakening $\omega_1$. This effectively causes the output neuron to spike only when input 1 does not spike, correctly implementing the NOT function.

After the training phase, the training layer is removed, and the spikes from the output neuron are decoded into a Bernoulli distribution. The output message is defined as
\begin{equation}
    \overrightarrow{\mu}_z(z) = \text{Ber}(z\,|\,p_z), 
        \quad \quad p_z = \rho/\tau
\end{equation}\label{eq:out}
where $\rho$ represents the total spike count observed at the output neuron, and $\tau$ is defined as the total number of samples drawn from the input messages. \bdv{Equations should be inserted in the sentence as if they are single nouns. So, there should be no semicolon here.}\sepideh{Done.} In this decoding process, $\mu_z(z)$ reflects the probability that the output neuron fires.

\begin{figure}
\centering
\includegraphics[width=1\columnwidth]{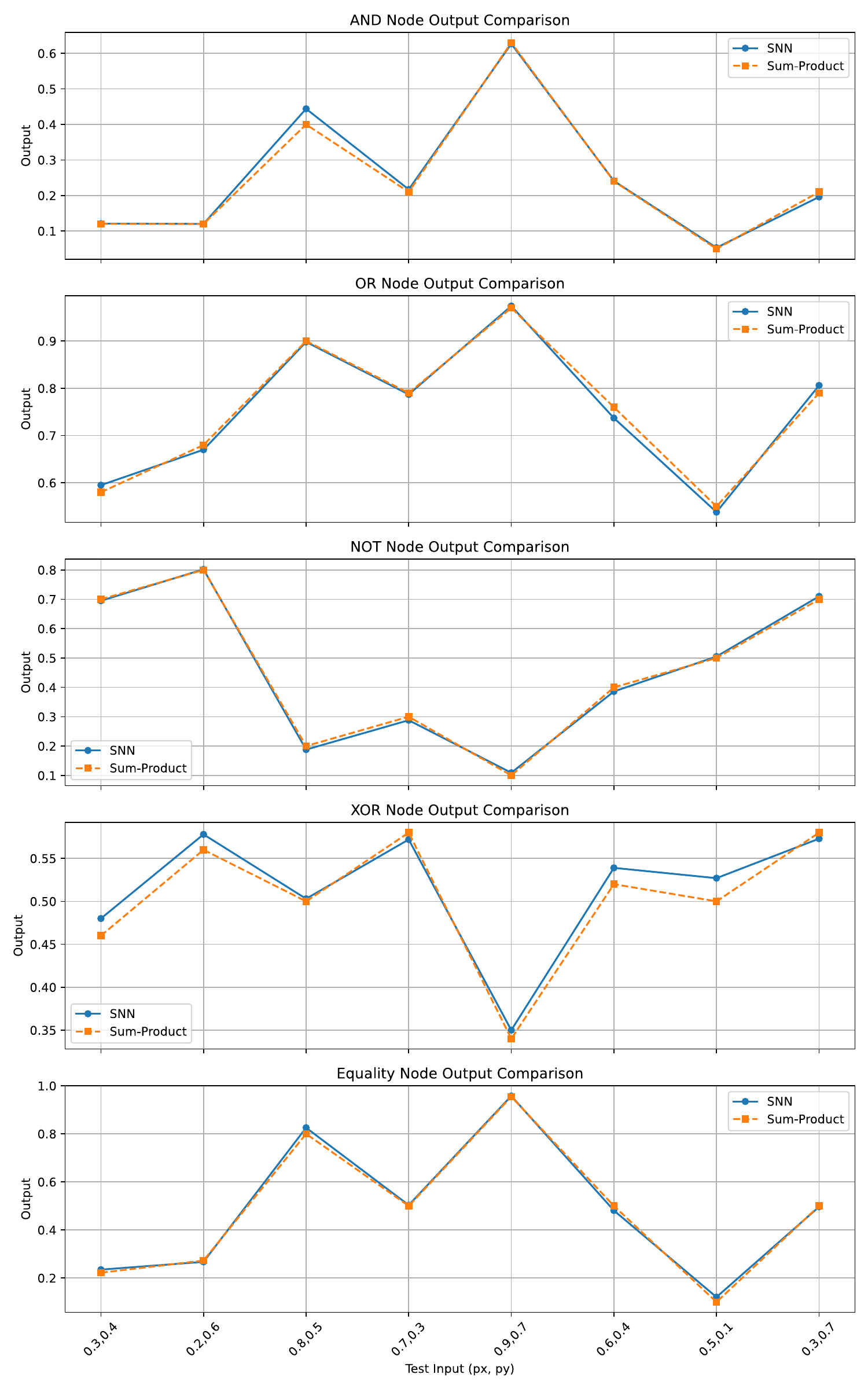}
\caption{Comparison of the results obtained from the proposed SNN-based nodes for passing Bernoulli messages with those produced by the sum-product algorithm. Validation was performed using eight random pairs of $p_x$ and $p_y$.}
\label{fig: result}
\end{figure}

% \begin{figure}
% \centering
% \includegraphics[width=\textwidth]{figures/logical_result.png}
% \caption{Comparison of the results obtained from the proposed SNN-based logic nodes for passing Bernoulli messages with those produced by the sum-product algorithm. Validation was performed using seven random pairs of $p_x$ and $p_y$.}
% \label{fig: result}
% \end{figure}

Figure~\ref{fig: result} compares these decoded results with the numerical outcomes obtained from the sum-product update rule defined in \eqref{eq:sum-product}, as summarized in Table~\ref{table:mp-logic}. The derivation details are provided in Appendix~\ref{A1}. As shown in Figure~\ref{fig: result}, the proposed spiking-based method closely matches the numerical results.

% \begin{figure}[tb]
%   \centering

%   \subfigure[AND Node]{
%     \includesvg[width=0.31\textwidth]{figures/synaptic_weights_AND.svg}
%     \label{fig:STDP_AND}
%   }
%   \hfill
%   \subfigure[OR Node]{
%     \includesvg[width=0.31\textwidth]{figures/synaptic_weights_OR.svg}
%     \label{fig:STDP_OR}
%   }
%   \hfill
%   \subfigure[NOT Node]{
%     \includesvg[width=0.31\textwidth]{figures/synaptic_weights_NOT.svg}
%     \label{fig:STDP_NOT}
%   }

%   \caption{Evolution of synaptic weights for AND, OR, and NOT nodes, from STDP-based training described in Section~\ref{sec: method}.. \bdv{In what experiment? Refer to the textual passage where you present this.}}
%   \label{fig:STDP_logic_nodes}
% \end{figure}

 \begin{table}[htbp]
\caption{Sum-product message update rules for the AND, OR, and NOT nodes, for given input messages $\overrightarrow\mu_x(x) = \mathcal{B}er(x | p_x)$ and $\overrightarrow\mu_y(y) = \mathcal{B}er(y | p_y)$. Detailed derivations are provided in Appendix~\ref{A1}. \bdv{Same as in other figures. LHS should indicate which message is $\overrightarrow\mu_z(z)$, etc.}\sepideh{Directions are added. The notation of messages matches the literature \cite{loeliger2004introduction}}
}
\centering
\setlength{\tabcolsep}{15pt}
\begin{tabular}{|c|c|}
\hline
 \textbf{\textit{Node function}}& \textbf{\textit{Update rule}} \\
\hline
  \begin{tabular}{c}\\ \begin{tikzpicture}
\node[smallbox] (g) {AND};
\node[below = 5 mm  of g] (down) {};
\node[left = 5 mm  of g] (left) {};
\node[right = 5 mm  of g] (right) {};

\draw[->] (left) -- (g)node[pos=0.5, above] {$X$};
\draw[->] (g) -- (right)node[pos=0.5, above] {$Z$};
\draw[<-] (g) -- (down)node[pos=0.5, right] {$Y$};
\end{tikzpicture} \end{tabular}& \begin{tabular}{c} 
  \normalsize $\overrightarrow\mu_z(z) = \mathcal{B}er(z \, | \, p_xp_y)$\\  \\  \end{tabular} \\
  \hline
  \begin{tabular}{c}\\ \begin{tikzpicture}
\node[smallbox] (g) {OR};
\node[below = 5 mm  of g] (down) {};
\node[left = 5 mm  of g] (left) {};
\node[right = 5 mm  of g] (right) {};

\draw[->] (left) -- (g)node[pos=0.5, above] {$X$};
\draw[->] (g) -- (right)node[pos=0.5, above] {$Z$};
\draw[<-] (g) -- (down)node[pos=0.5, right] {$Y$};
\end{tikzpicture}\end{tabular} &  \begin{tabular}{c} \normalsize $ \overrightarrow\mu_z(z) = \mathcal{B}er(z \, | \, p_x-p_x p_y+p_y)\,$\\ \\  \end{tabular}\\
\hline
    \begin{tabular}{c}\\ \begin{tikzpicture}
\node[smallbox] (g) {NOT};
\node[left = 5 mm  of g] (left) {};
\node[right = 5 mm  of g] (right) {};

\draw[->] (left) -- (g)node[pos=0.5, above] {$X$};
\draw[->] (g) -- (right)node[pos=0.5, above] {$Z$};
\end{tikzpicture}\end{tabular} &  \begin{tabular}{c} \normalsize $ \overrightarrow\mu_z(z) = \mathcal{B}er(z \, | \, 1 - p_x)\,$\\ \\  \end{tabular}\\
\hline
\end{tabular}
\vspace{-4mm}
\label{table:mp-logic}
\end{table}

The proposed networks can be used to implement more complex logical operations. As an example, we construct the XOR operation using combinations of the basic logical networks; its architecture is shown in Figure \ref{fig: xor}. 
In Figure~\ref{fig: result}, the output of the spiking-based XOR implementation is compared with the numerical results obtained from the sum-product update rule, as presented in Table~\ref{table:mp-b}.

\begin{figure}[tb]
\centerline{{\begin{tikzpicture}[
roundnode/.style={circle, draw=black!60, very thick, minimum size=7mm},
]
\node[roundnode] (AND) {\textit{AND}};
\node[roundnode,below left = 4 mm and 20 mm of AND] (or) {\textit{OR}};

\node[right = 14 mm of AND] (out){\textit{$Z$}};
\node[right = 9 mm  of AND, rotate=90, anchor=south] (dec) {\textit{Decoder}};
\node[fit=(dec), draw, inner sep = .0 mm] (box_dec) {};

\node[roundnode, left = 20 mm  of AND] (NOT) {\textit{NOT}};
\node[roundnode,left = 8 mm of NOT] (and1) {\textit{AND}};
\node[right = 1 mm  of left] (w13) {};
\node[right = 1 mm  of or] (w13) {};
\node[above = 3 mm  of NOT] (title) {\Large\textit{XOR}};

\node[left = 20 mm of and1] (s1){\textit{$X$}};
\node[left = 10 mm  of and1, rotate=90, anchor=south] (en_x) {\textit{Encoder}};
\node[fit=(en_x), draw, inner sep = .0 mm] (box_en_x) {};

\node[below = 15 mm of or] (s2){\textit{$Y$}};
\node[below = 8 mm  of or] (en_y) {\textit{Encoder}};
\node[fit=(en_y), draw, inner sep = .0 mm] (box_en_y) {};

\node[below = 2.5 mm of or] (space){};
\node[below = 16.4 mm of and1] (space2){};
\node[left = 4 mm of and1] (space3){};
\node[left = 25 mm of or] (space4){};

\draw[<-] (AND)  |- (or);
\draw[->] (and1) -- (NOT);
\draw[->] (AND) -- (box_dec);
\draw[->] (box_dec) -- (out);

\draw[->] (s2) -- (box_en_y);
\draw[->] (box_en_y)--(or);

\draw[->] (s1) -- (box_en_x);
\draw[->] (box_en_x)--(and1);

\draw[<-] (AND) -- (NOT);
\draw[-] (-5,-2) -- (-2.72,-2);
\draw[->] (-5,-2) -- (and1);
\draw[->] (-5.8,-1.15) -- (or);
\draw[-] (-5.8,-1.15) -- (-5.8,0);

\node[fit=(AND)(space)(space3)(NOT)(or)(and1), draw, inner sep = 2.8 mm] (box) {};

\end{tikzpicture}}}
\caption{SNN architecture for simulating the sum-product update rule of XOR factor node as defined in Table~\ref{table:mp-b}. The component networks reused from the AND, OR, and NOT implementations are shown in Figure~\ref{fig: logical_gates1}, and the same encoding and decoding methods are applied. \bdv{What type of graph is this? How should I read the $X$ and $Y$ variables in the edges? These edges have arrows on both sides. If you draw an FFG, please stick to proper FFG conventions. If you don't, then write how to read the graph.}\sepideh{I modified the text a little.}}
\label{fig: xor}
\end{figure}

\begin{wraptable}[11]{r}{0.5\textwidth}
\vspace{-1.5em}
\centering
\caption{Equality Constraint. The table presents the expected output spike behavior of an equality node. %given different combinations of input spikes.%\bdv{Insufficient caption (explain what this is and context). In particular, the "not defined" output is concerning.}
}
\label{table:equality}
\setlength{\tabcolsep}{12pt}
\begin{tabular}{|c|c|}
\hline
\textbf{$S_1\quad S_2$} & \textbf{Equality} \\
\hline
$0 \quad 0$ & 0 \\
$0 \quad 1$ & Not Defined \\
$1 \quad 0$ & Not Defined \\
$1 \quad 1$ & 1 \\
\hline
\end{tabular}
\vspace{15pt}
\end{wraptable}
The final operation we address in this paper is the equality constraint, a fundamental factor node in the sum-product algorithm. Although it is not a logical operation in the traditional sense, it can still be simulated using the proposed spiking networks. We expect a behavior consistent with the values reported in Table~\ref{table:equality} for the equality constraint. As shown in the table, the total number of output spikes, denoted by $\rho$ in~\eqref{eq:out}, closely resembles the result of an AND operation applied to the input spike trains. To determine an appropriate value for $\tau$ in~\eqref{eq:out} under this constraint, cases labeled as 'Not Defined' are excluded from the total number of samples. Interestingly, these cases align with the spike count of the XOR logical operation.
\bdv{Huh? You "route" the spikes somewhere else? Do you have a proper FFG-based process for this or do you do this through a backdoor? this is concerning.}\sepideh{Text modified.} The final outcome is compared in Figure~\ref{fig: result} with the numerical sum-product result, obtained using the update rules presented in Table~\ref{table:mp-b}.

\section{Example}

\begin{figure*}[htbp] 
\centering
\begin{minipage}[c]{0.48\textwidth}
\centering
\begin{tikzpicture}
\node[smallbox] (g) {$=$};
\node[above left = 1 mm and 3mm  of g] (y) {};
\node[smallbox, left = 10 mm  of g] (xor) {XOR};
\node[smallbox, below = 10 mm  of g] (x_3) {$\mathcal{B}er$};
\node[smallbox, below = 10 mm  of xor] (x_2) {$\mathcal{B}er$};
\node[smallbox, right = 10 mm  of x_3] (x_4) {$\mathcal{B}er$};
\node[smallbox, left = 10 mm  of x_2] (x_1) {$\mathcal{B}er$};
\node[clamped, below = 5 mm  of x_4] (z_4) {};
\node[clamped, below = 5 mm  of x_3] (z_3) {};
\node[clamped, below = 5 mm  of x_2] (z_2) {};
\node[clamped, below = 5 mm  of x_1] (z_1) {};
\node[below = 1 mm  of z_1] (zz_1) {$y_1$};
\node[below = 1 mm  of z_2] (zz_2) {$y_2$};
\node[below = 1 mm  of z_3] (zz_3) {$y_3$};
\node[below = 1 mm  of z_4] (zz_4) {$y_4$};

% \draw[-] (xor) -- (g)node[teal,pos=0.5, above] {$\rightarrow $}node[teal,pos=0.5, below] {$\leftarrow$};
% \draw[-] (x_2) -- (xor)node[orange,pos=0.5, right] {$\uparrow$}node[ pos=0.5, right] {$\quad x_2$}node[teal,pos=0.5, left] {$\downarrow$};
% \draw[-] (g) -- (x_3)node[orange, pos=0.5, right] {$\uparrow$}node[ pos=0.5, right] {$\quad x_3$}node[teal, pos=0.5, left] {$\downarrow$};
% \draw[-] (x_4) |- (g)node[orange, pos=0.2, right] {$\uparrow$}node[ pos=0.2, right] {$\quad x_4$}node[teal, pos=0.2, left] {$\downarrow$};
% \draw[-] (x_1) |- (xor)node[orange, pos=0.2, right] {$\uparrow$}node[ pos=0.2, right] {$\quad x_1$}node[teal, pos=0.2, left] {$\downarrow$};

\draw[->] (xor) -> (g)node[pos=0.5, above] {$z$};
\draw[->] (x_2) -- (xor)node[ pos=0.5, right] {$x_2$};
\draw[<-] (g) -- (x_3)node[ pos=0.5, right] {$x_3$};
\draw[->] (x_4) |- (g)node[ pos=0.2, right] {$ x_4$};
\draw[->] (x_1) |- (xor)node[ pos=0.2, right] {$x_1$};

\draw[-] (x_1) -- (z_1);
\draw[-] (x_2) -- (z_2);
\draw[-] (x_3) -- (z_3);
\draw[-] (x_4) -- (z_4);

\end{tikzpicture}
% \label{fig:example}
\end{minipage}
\hfill
\begin{minipage}[c]{0.48\textwidth}
\centering
% \captionof{table}{The unreliable channel example corresponding to Figure~\ref{fig:example}. \bdv{Fig\ref{fig:example} does not indicate any forward or backward direction, so it is unclear in which direction forward messages flow.}\sepideh{Done.}}
% \label{table:example1}
\setlength{\tabcolsep}{7pt}
\renewcommand{\arraystretch}{1.1}
\begin{tabular}{ c|c|c}
\hline 
\textbf{\textit{Message}} & \textbf{\textit{SP ~\cite{loeliger2004introduction}}} & \textbf{\textit{SNNs}} \\
\hline 
$\overrightarrow{\mu}_z(z)$ & $\mathcal{B}er(0.180)$ & $\mathcal{B}er(0.174)$ \\
$\overleftarrow{\mu}_z(z)$ & $\mathcal{B}er(0.500)$ & $\mathcal{B}er(0.488)$ \\
$\overleftarrow{\mu}_{x_1}(x_1)$ & $\mathcal{B}er(0.500)$ & $\mathcal{B}er(0.526)$ \\
$\overleftarrow{\mu}_{x_2}(x_2)$ & $\mathcal{B}er(0.500)$ & $\mathcal{B}er(0.526)$ \\
$\overleftarrow{\mu}_{x_3}(x_3)$ & $\mathcal{B}er(0.024)$ & $\mathcal{B}er(0.022)$ \\
$\overleftarrow{\mu}_{x_4}(x_4)$ & $\mathcal{B}er(0.664)$ & $\mathcal{B}er(0.657)$ \\
% \hline
\end{tabular}
\end{minipage}
\caption{(Left) The FFG corresponding to the unreliable binary channel example. (Right) The sum-product calculated messages versus messages produced by the SNNs.}
\label{fig:example}
\vspace{-10pt}
\end{figure*}
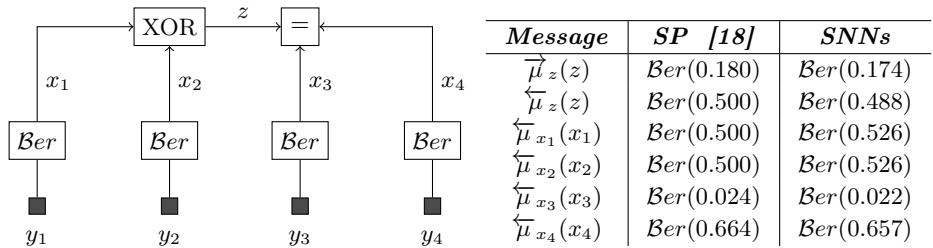

In this section, we evaluate the inference capability of the proposed spiking network model using an example based on an unreliable binary communication channel. This signal transmission scenario has been used in prior work~\cite{loeliger2004introduction}~\cite{steimer2009belief} as a benchmark for evaluating Bernoulli message-passing algorithms. \bdv{For some reason, I often see in your papers that there is no space between a word and the citation. Always insert a space.}\sepideh{Done.} Consider a simple binary code  $C = \{(0,0,0,0),\ (0,1,1,1),\ (1,0,1,1),\ (1,1,0,0)\} $ illustrated as a FFG in Figure~\ref{fig:example} (left).
Each bit of a binary codeword $X = (x_1, x_2, x_3, x_4)$
is transmitted through an unreliable communication channel with a known bit-flip (cross-over) probability of \( \varepsilon = 0.1 \).
The generative model is given by
\begin{align}
    f(x_1, &\dots x_4, z  \, | \, y_1, \dots y_4) \propto  \, f_{xor}(x_1,x_2,z) \, f_=(x_3,x_4,z) \, \prod_{i=1}^{4}p(y_i \, | \, x_i).\nonumber
\end{align} 
where $z$ is a latent variable and the $y_i$ are noisy observations. Given the observed values $(y_1, y_2, y_3, y_4) = \left(0, 0 , 1, 0\right)$, the corresponding Bernoulli messages can be obtain as 
\begin{align}
     &\left(\overrightarrow{\mu}_{x_1}(x_1), \overrightarrow{\mu}_{x_2}(x_2), \overrightarrow{\mu}_{x_3}(x_3), \overrightarrow{\mu}_{x_4} (x_4)\right) = \left(\mathcal{B}er(0.1),\mathcal{B}er(0.1),\mathcal{B}er(0.9),\mathcal{B}er(0.1) \right) \nonumber .
    \label{eq:example-messages}
\end{align} as detailed in Appendix~\ref{A2}.
\bdv{Left-hand side of the model is conditioned on $\hat{y}_i$, while right-hand side has $\hat{y}_i$ as a dependent variable. This cannot be true, at the very least, use a $\propto$ sign, but review first.}\sepideh{Done. This model matches the model in \cite{steimer2009belief}}
\bdv{The last sentence is not a proper sentence, which makes me worry that you have not used proper spelling corrective software.}\sepideh{fixed.}

The goal is to compute the marginal posterior 
%probabilities  $P(x_i \mid \hat{y}_1, \hat{y}_2, \hat{y}_3, \hat{y}_4)$
for each bit \( x_i \). The table in Figure \ref{fig:example} (right) compares all the messages produced by our SNN-based model with the ground-truth messages passed along each edge of the FFG. The results show that our method closely approximates the expected message values. Finally, the marginal posteriors are obtained via the normalized product  
\begin{align}\overrightarrow{\mu}_{x_i}(x_i) \times \overleftarrow{\mu}_{x_i}(x_i) = {Ber(0.109), Ber(0.109), Ber(0.168), Ber(0.175)}, 
\end{align}
which closely match the messages produced by the sum-product rule $Ber(0.1)$, $Ber(0.1)$, $Ber(0.18)$, $Ber(0.18)$ \cite{loeliger2004introduction}.
\bdv{mach? Please use spelling checkers. }\sepideh{Fixed.}

% \begin{figure}[tb]
% \centerline{\input{graphs/channel-example.tikz}}
% \caption{The FFG corresponding to the unreliable binary channel example.}
% \label{fig: example}
% \end{figure}

% \begin{table}[htbp]
% \caption{The unreliable channel example \ref{fig: example}.
% }
% \begin{center}
% \begin{tabular}{|c|c|c|}
% \hline 
%  \textbf{\textit{Message}}& \textbf{\textit{Ground truth~\cite{loeliger2004introduction}}} & \textbf{\textit{Our result}} \\
%  \hline 
%   $\overrightarrow{\mu}_z(z)$ &  $\mathcal{B}er(0.180) $& $\mathcal{B}er(0.174)$\\
%   $\overleftarrow{\mu}_z(z)$ & $\mathcal{B}er(0.500)$  & $\mathcal{B}er(0.488)$\\
%   $\overleftarrow{\mu}_{x_1}(x_1)$ & $\mathcal{B}er(0.500)$  & $\mathcal{B}er(0.526)$\\
%   $\overleftarrow{\mu}_{x_2}(x_2)$ & $\mathcal{B}er(0.500) $ & $\mathcal{B}er(0.526)$\\
%   $\overleftarrow{\mu}_{x_3}(x_3)$ & $\mathcal{B}er(0.024)$  & $\mathcal{B}er(0.022)$\\
%   $\overleftarrow{\mu}_{x_4}(x_4)$ & $\mathcal{B}er(0.664)$ & $\mathcal{B}er(0.657)$\\
% \hline
% \end{tabular}
% \label{table:example1}
% \end{center}
% \end{table}

% \begin{figure*}[htbp]
% \centering

% \begin{minipage}[c]{0.48\textwidth}
% \centering
% \input{graphs/channel-example.tikz}
% \caption{The FFG corresponding to the unreliable binary channel example.}
% \label{fig:example}
% \end{minipage}
% \hfill
% \begin{minipage}[c]{0.48\textwidth}
% \centering
% \captionof{table}{The unreliable channel example corresponding to Figure~\ref{fig:example}. \bdv{Fig\ref{fig:example} does not indicate any forward or backward direction, so it is unclear in which direction forward messages flow.}\sepideh{Done.}}
% \label{table:example1}
% \begin{tabular}{|c|c|c|}
% \hline 
% \textbf{\textit{Message}} & \textbf{\textit{Ground truth~\cite{loeliger2004introduction}}} & \textbf{\textit{Our result}} \\
% \hline 
% $\overrightarrow{\mu}_z(z)$ & $\mathcal{B}er(0.180)$ & $\mathcal{B}er(0.174)$ \\
% $\overleftarrow{\mu}_z(z)$ & $\mathcal{B}er(0.500)$ & $\mathcal{B}er(0.488)$ \\
% $\overleftarrow{\mu}_{x_1}(x_1)$ & $\mathcal{B}er(0.500)$ & $\mathcal{B}er(0.526)$ \\
% $\overleftarrow{\mu}_{x_2}(x_2)$ & $\mathcal{B}er(0.500)$ & $\mathcal{B}er(0.526)$ \\
% $\overleftarrow{\mu}_{x_3}(x_3)$ & $\mathcal{B}er(0.024)$ & $\mathcal{B}er(0.022)$ \\
% $\overleftarrow{\mu}_{x_4}(x_4)$ & $\mathcal{B}er(0.664)$ & $\mathcal{B}er(0.657)$ \\
% \hline
% \end{tabular}
% \end{minipage}
% \end{figure*}

\section{Conclusions}

We presented a biologically plausible network of spiking neurons trained using STDP to perform message-passing for Bernoulli messages. This work represents a potential step toward bridging the gap between theoretical models of brain function, such as the Bayesian Brain hypothesis and the FEP, and the spiking behavior of biological neurons.

\begin{credits}
\subsubsection{\ackname} 
This work was carried out in the context of the BayesBrain project. We gratefully acknowledge financial support from the Eindhoven Artificial Intelligence Systems Institute (EAISI) at TU Eindhoven. 
% To be added in the final version.
\bdv{If the submission is double-blind, please replace the acknowledgment statement by "will be added later."}
\sepideh{Done.}

\subsubsection{\discintname}
The authors declare no conflict of interest.
\end{credits}

\appendix
\section{Derivation of Sum-Product Update Rules}\label{A1}

We can derive the forward sum-product message $ \overrightarrow{\mu}_z\left(z\right)$ according to \eqref{eq:sum-product}, given the input messages $\overrightarrow{\mu}_x\left(x\right) = \mathcal{B}er\left(x| p_x\right)$ and $\overrightarrow{\mu}_y\left(y\right) = \mathcal{B}er\left(y | p_y\right)$.  
\begin{align*}
    \overrightarrow{\mu}_z(z) &= \sum_{x} \sum_{y} \overrightarrow{\mu}_x\big(x\big)\,\overrightarrow{\mu}_y\big(y\big)\,f\big(z, x, y\big) \\[1pt]
    % &= \sum_{x} \sum_{y} \mathcal{B}er\!\big(x | p_{x}\big)\,\mathcal{B}er\!\big(y | p_{y}\big)\,f\big(z, x, y\big)\\[1pt]
    &= \mathcal{B}er\!\big(0 | p_{x}\big)\,\mathcal{B}er\!\big(0 | p_{y}\big))f\big(z, 0, 0\big) + \mathcal{B}er\!\big(1 | p_{x}\big)\,\mathcal{B}er\!\big(0 | p_{y}\big)f\big(z, 1, 0\big)\\[2pt]
    &\quad + \mathcal{B}er\!\big(0 | p_{x}\big)\,\mathcal{B}er\!\big(1 | p_{y}\big)f\big(z, 0, 1\big) +\mathcal{B}er\!\big(1 | p_{x}\big)\,\mathcal{B}er\!\big(1 | p_{y}\big)f\big(z, 1, 1\big)\\[2pt]
    &= \big(1 - p_{x}\big)\big(1 - p_{y}\big)f\big(z, 0, 0\big) + p_{x}\big(1 - p_{y}\big)f\big(z,1, 0\big)\\[2pt]
    &\quad + \big(1 - p_{x}\big)p_{y} f\big(z, 0, 1\big) + p_{x} p_{y} f\big(z, 1, 1\big)\,.
\end{align*}

Using the truth table, we can substitute $z$ and evaluate the terms for different operations. As an example, the AND operation is 
\begin{align*}
    \overrightarrow{\mu}_z(z) &= \begin{cases} p_xp_y  &\text{ if } z=1\\
    \left( 1 - p_x\right)\left( 1 - p_y\right) + p_x \left( 1-p_y\right) + \left( 1-p_x\right)p_y &\text{ if } z=0 \end{cases}
    \\
    &= 
    \quad
    \begin{cases} p_x p_y&\text{ if } z=1\\
    1 - p_x p_y &\text{ if } z=0 \end{cases} \quad=\quad \mathcal{B}er\left( z | p_x p_y\right)\,.
\end{align*}
Similarly, we have the following computations for the OR factor node
\begin{align*}
    \overrightarrow{\mu}_z(z) &= \begin{cases} p_xp_y + p_x \left( 1-p_y\right) + \left( 1-p_x\right)p_y  &\text{ if } z=1\\
    \left( 1 - p_x\right)\left( 1 - p_y\right) &\text{ if } z=0 \end{cases}
    \\
    &= 
    \quad
    \begin{cases} p_x + p_y -p_x p_y&\text{ if } z=1\\
    1 - p_x - p_y + p_x p_y &\text{ if } z=0 \end{cases} \quad=\quad \mathcal{B}er\left( z | p_x + p_y -p_x p_y\right)\,.
\end{align*}

\section{Derivation of Messages in the Example}\label{A2}
Under the bit-flip channel model, the likelihood is given by 
\begin{align*}
    p\left(y_i | x_i\right) &= \begin{cases}\left(1 - \varepsilon\right)x_i + \varepsilon\left(1-x_i\right)  &\text{ if } y_i=1\\
    \varepsilon x_i + \left( 1 - \varepsilon\right)\left( 1 - x_i\right) &\text{ if } y_i=0 \end{cases} \ = \ \mathcal{B}er\left( y_i |\left(1 - \varepsilon\right)x_i + \varepsilon\left(1-x_i\right) \right)\,.
\end{align*}
According to the sum-product rule, the messages can be obtained by
\begin{align*}
    \overrightarrow\mu_{x_i}(x_i) = &\sum_{y_i \in \{0,1\}} \delta(y_i - \hat y_i) \mathcal{B}er\left( y_i |\left(1 - \varepsilon\right)x_i + \varepsilon\left(1-x_i\right) \right)  \\
    &= \begin{cases}\left(1 - \varepsilon\right)x_i + \varepsilon\left(1-x_i\right)  &\text{ if } \hat{y}_i=1\\
    \varepsilon x_i + \left( 1 - \varepsilon\right)\left( 1 - x_i\right) &\text{ if } \hat{y}_i=0 \end{cases} \\
    &= \begin{cases}\mathcal{B}er\left( x_i |1-\varepsilon\right)  &\text{ if } \hat{y}_i=1\\
    \mathcal{B}er\left( x_i |\varepsilon\right)  &\text{ if } \hat{y}_i=0 \end{cases} .
\end{align*}


\bdv{Perhaps add references \url{https://arxiv.org/abs/2410.19315} and \url{https://arxiv.org/abs/2505.17962}? } 
\bibliographystyle{splncs04}
\bibliography{content/references.bib}

\end{document}